\documentclass[prl,twocolumn,superscriptaddress,showpacs]{revtex4}
\usepackage{graphicx,amsmath,amssymb,bm,dcolumn}

\newcommand{\be}{\begin{equation}}	
\newcommand{\ee}{\end{equation}}

\newcommand{\fm}{\, \text{fm}}

\newcommand{\fmiq}{\, \text{fm}^{-3}}
\newcommand{\mev}{\, \text{MeV}}
\newcommand{\gcmiq}{\, \text{g} \, \text{cm}^{-3}}
\newcommand{\km}{\, \text{km}}

\begin{document}

\title{Constraints on neutron star radii based on chiral effective
field theory interactions}

\author{K.\ Hebeler}
\email[E-mail:~]{hebeler@triumf.ca}
\affiliation{TRIUMF, 4004 Wesbrook Mall, Vancouver, BC, V6T 2A3, Canada}
\author{J.\ M.\ Lattimer}
\email[E-mail:~]{lattimer@astro.sunysb.edu}
\affiliation{Department of Physics and Astronomy, Stony Brook
University, Stony Brook, NY 11794-3800, USA}
\author{C.\ J.\ Pethick}
\email[E-mail:~]{pethick@nbi.dk}
\affiliation{The Niels Bohr International Academy, The Niels Bohr Institute,
Blegdamsvej 17, DK-2100 Copenhagen \O, Denmark}
\affiliation{NORDITA, Roslagstullsbacken 21, SE-10691 Stockholm, Sweden}
\author{A.\ Schwenk}
\email[E-mail:~]{schwenk@physik.tu-darmstadt.de}
\affiliation{ExtreMe Matter Institute EMMI, GSI Helmholtzzentrum f\"ur
Schwerionenforschung GmbH, 64291 Darmstadt, Germany}
\affiliation{Institut f\"ur Kernphysik, Technische Universit\"at
Darmstadt, 64289 Darmstadt, Germany}
\affiliation{TRIUMF, 4004 Wesbrook Mall, Vancouver, BC, V6T 2A3, Canada}

\begin{abstract}
We show that microscopic calculations based on chiral effective
field theory interactions constrain the properties of neutron-rich
matter below nuclear densities to a much higher degree than is
reflected in commonly used equations of state. Combined with
observed neutron star masses, our results lead to a radius $R = 9.7
- 13.9 \km$ for a $1.4 \, M_\odot$ star, where the theoretical
range is due, in about equal amounts, to uncertainties in many-body
forces and to the extrapolation to high densities.
\end{abstract}


\maketitle

{\it Introduction.--} With the advances in observational capabilities,
it is becoming possible to obtain direct evidence for the size of
neutron stars~\cite{nsreview}. Sources of information include
measurements of optical radiation from nearby isolated neutron stars
whose distances are known from parallax determinations, and
observations of thermonuclear flares on the surfaces of neutron
stars~\cite{Ozel2010,Steiner2010}. Neutron star
seismology~\cite{Samuelsson2007}, pulse profiles in X-ray
pulsars~\cite{Leahy2009} and moment of inertia
measurements~\cite{Lattimer05} are additional sources. In the near
future, one further expects gravitational wave signals from collapsing
stars and merging binary neutron stars to provide information about
the equation of state of dense matter~\cite{Ferrari2010}.

In nuclear physics, recent developments of effective field theory
(EFT) and the renormalization group (RG) for nuclear forces enable
controlled calculations of matter at nuclear
densities~\cite{RMP,PPNP,nucmatt}. In this framework, it is possible
to estimate the theoretical uncertainties due to neglected many-body
forces and from an incomplete many-body calculation. In this Letter,
we show that such calculations of the equation of state
(EOS) below nuclear densities constrain the properties of dense
matter, and the radii of typical neutron stars, to a much higher
degree than is reflected in current neutron star modeling.

\begin{figure*}[t]
\begin{center}
\includegraphics[scale=0.75,clip=]{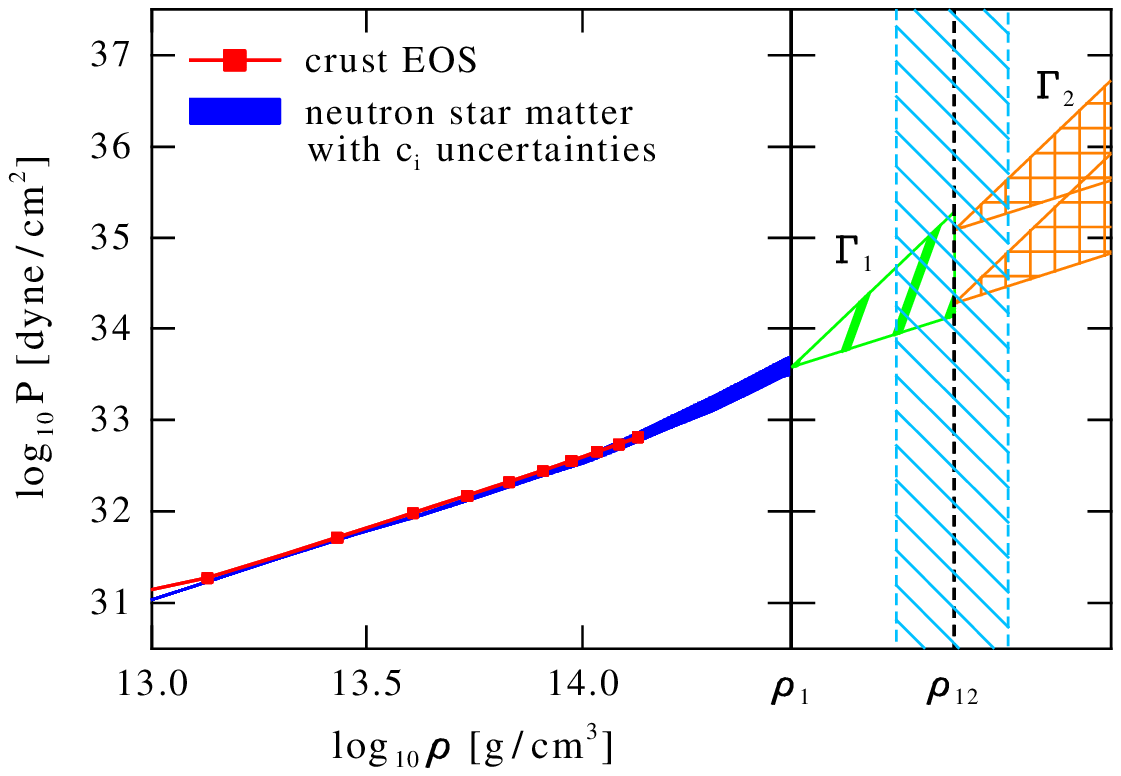}
\hspace*{5mm}
\includegraphics[scale=0.75,clip=]{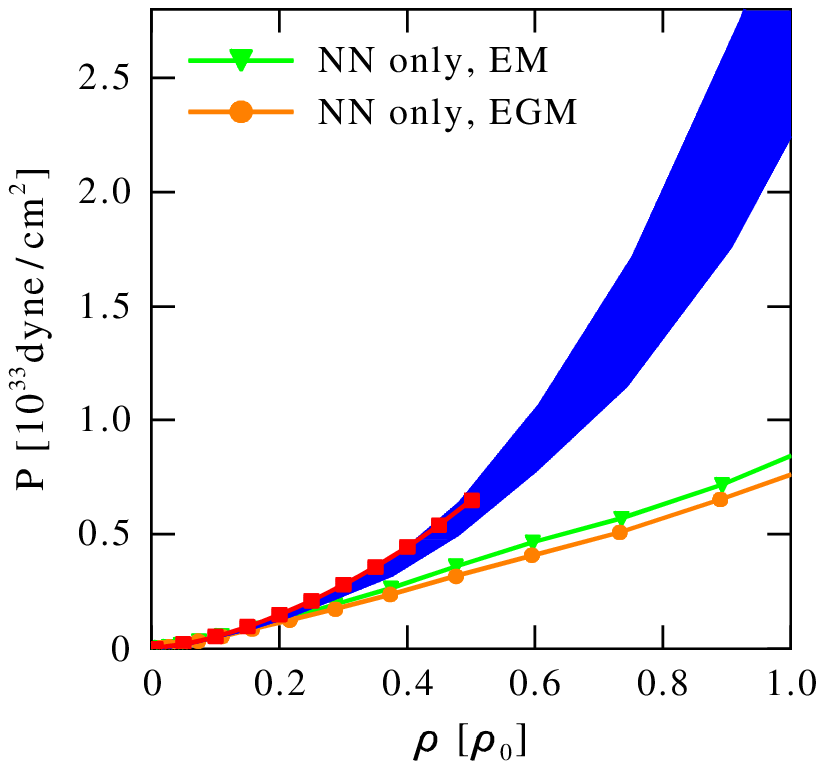}
\end{center}
\vspace*{-4mm}
\caption{Pressure of neutron star matter based on chiral low-momentum
interactions for densities $\rho < \rho_1$ (corresponding to a neutron
density $\rho_{1,n} = 1.1 \rho_0$). The band estimates the theoretical
uncertainties from many-body forces and from an incomplete many-body
calculation. At low densities, the results are compared to a standard
crust EOS~\cite{crust}, where the right panel demonstrates the
importance of 3N forces. The extension to higher densities using
piecewise polytropes (as explained in the text) is illustrated
schematically in the left panel.}
\vspace*{-4mm}
\label{prho}
\end{figure*}

{\it Neutron matter below nuclear densities.--} Our studies are based
on microscopic calculations of neutron matter using chiral
low-momentum two-nucleon (NN) and three-nucleon (3N)
interactions~\cite{nm}. The largest uncertainty for the neutron matter
energy arises from the couplings $c_3$ and (to a smaller extent) $c_1$
that determine the leading two-pion-exchange three-body forces among
neutrons in chiral EFT. We improve the range of $c_3$ compared to
Ref.~\cite{nm} by taking $c_3$ values from extractions based on the
same couplings in the subleading two-pion-exchange NN interaction,
$c_3 = -(3.2 - 4.8) \, {\rm GeV}^{-1}$~\cite{EM,EGM,Rentmeester}. In
addition, we include a shift $\delta c_3 = 1.0 \, {\rm GeV}^{-1}$ to
take into account contributions at the next order for 3N
forces~\cite{RMP}. In the following we therefore use $c_3 = -(2.2 -
4.8) \, {\rm GeV}^{-1}$ and $c_1 = - (0.7 - 1.4) \, {\rm GeV}^{-1}$
with a smooth $n_{\rm exp}^{\rm 3N} = 4$ regulator. The neutron matter
calculations details are as in Ref.~\cite{nm}, which suggested that
the energy is perturbative at nuclear densities. Using only NN
interactions, we obtain a neutron matter energy per particle $E^{\rm
NN}_n(\rho_0)/N = 10.4 \mev$ at saturation density $\rho_0 = 2.7
\times 10^{14} \gcmiq$. (We define the density $\rho$ as the nucleon
mass times the baryon density.) The inclusion of 3N forces leads to
$E_n(\rho_0)/N = 16.3 \pm 2.2 \mev$, dominated by the repulsive $c_3$
contribution and the associated uncertainty.  The 3N contribution of
$\approx 6 \mev$ is to be compared to the NN potential energy $\approx
-26 \mev$ (the kinetic energy is $3/5 \varepsilon_{\rm F}(\rho_0)
\approx 36 \mev$), and also the 3N uncertainty of $\approx 2 \mev$ is
consistent with the contributions of higher-order 3N forces (given an
approximate expansion parameter in chiral EFT of $\sim 1/3$ at these
momenta). Other microscopic calculations lie within our theoretical
uncertainties~\cite{nm}, and at low densities $\rho \sim \rho_0/10$,
the results are also consistent with calculations for resonant Fermi
gases including effective range contributions~\cite{coldatoms}.

{\it Neutron star matter.--} We extend our results to matter in beta
equilibrium using the parametrization: 
\be
\frac{E(\rho,x)}{A} = \frac{E_n(\rho)}{N} - 4 x (1-x) S_2(\rho) +
\frac{3 x \hbar c}{4} \, (3 \pi^2 x \rho/m)^{1/3} \,,
\label{energy}
\ee
where $E_n(\rho)/N$ is given by our neutron matter results, $m$
is the nucleon mass, and $x$ the proton fraction. The last terms in
Eq.~(\ref{energy}) incorporate the contributions from protons through
the symmetry energy $S_2(\rho)$ and from electrons~\cite{LP}. The
proton fraction in beta equilibrium is given approximately by $x(\rho)
\approx [4 S_2(\rho)/(\hbar c)]^3/(3 \pi^2 \rho/m)$, and for 
$S_2(\rho_0) \approx 30 \mev$,  $x(\rho_0) \approx 5 \%$. The energy
difference between neutron matter and matter in beta
equilibrium is $-1.1 \mev$, and this
will have only a minor
impact on our final results. We extract $S_2(\rho)$ for nuclear
densities using empirical saturation properties, \be S_2(\rho) =
\frac{E_n(\rho)}{N} + a_V - \frac{K}{18 \rho_0^2} \, (\rho - \rho_0)^2
\,,
\label{symfit}
\ee
with binding energy $a_V = 16 \mev$ and incompressibility $K = 230
\mev$ (which are within theoretical uncertainties of the nuclear
matter calculations of Ref.~\cite{nucmatt}). To include the symmetry
energy in Eq.~(\ref{energy}), we use the Ansatz $S_2(\rho) =
\overline{S}_2 (\rho/\rho_0)^{\gamma}$ and fit $\overline{S}_2 \equiv
S_2(\rho_0)$ and $\gamma$ to our neutron matter results. The fit has a
relative uncertainty of $< 5\%$ for densities $\rho_0/8 < \rho < 
\rho_1 = 3.0 \times 10^{14} \gcmiq$ ($\rho_1$ corresponds to a neutron
density $\rho_{1,n} = 1.1 \rho_0$). We obtain the following symmetry energy
parameters and proton fractions:
\begin{center}
\begin{tabular}{c|c||c|c|c}
\: $c_1 \, [{\rm GeV}^{-1}]$ \: & \: $c_3 \, [{\rm GeV}^{-1}]$ \: &
\: $\overline{S}_2 \, [{\rm MeV}]$ \: & $\gamma$ & $x(\rho_0)$ 
\\ \hline\hline 
\quad $-0.7$ \quad\quad & \quad $-2.2$ \quad\quad & \quad $30.1$
\quad\: & \: $0.5$ \quad & \: $4.8\%$ \quad \\ 
\quad $-1.4$ \quad\quad & \quad $-4.8$ \quad\quad & \quad $34.4$
\quad\: & \: $0.6$ \quad & \: $7.2\%$ \quad \\ \hline 
\multicolumn{2}{c||}{NN-only EM} & \quad $26.5$ \quad\: & \: $0.4$
\quad & \: $3.3\%$ \quad \\
\multicolumn{2}{c||}{NN-only EGM} & \quad $25.6$ \quad\: & \: $0.4$
\quad & \: $2.9\%$ \quad
\end{tabular}
\end{center}

The resulting pressure of neutron star matter is shown in
Fig.~\ref{prho} for densities $\rho < \rho_1$, where the band is
dominated by the uncertainty in $c_3$. The comparison of these
parameter-free calculations to a standard crust EOS~\cite{crust} shows
good agreement to low densities $\rho \gtrsim \rho_0/10$ within the
theoretical uncertainties.
In addition, the right panel of
Fig.~\ref{prho} demonstrates the importance of 3N forces. The pressure
obtained from low-momentum NN interactions only, based on the
RG-evolved N$^3$LO potentials of Entem and Machleidt
(EM)~\cite{EM} or Epelbaum {\it et al.} (EGM)~\cite{EGM}, differ
significantly from the crust EOS at $\rho_0/2$.

{\it Neutron stars.--} The structure of non-rotating
neutron stars without magnetic fields is determined by solving the
Tolman-Oppenheimer-Volkov (TOV) equations.
Because the central densities reach values higher than $\rho_1$, we
need to extend the uncertainty band for the pressure of neutron star
matter beyond $\rho_1$. To this end, we introduce a transition density
$\rho_{12}$ that separates two higher-density regions, and describe
the pressure by piecewise polytropes, $P(\rho) = \kappa_1
\rho^{\Gamma_1}$ for $\rho_1 < \rho < \rho_{12}$, and $P(\rho) =
\kappa_2 \rho^{\Gamma_2}$ for $\rho > \rho_{12}$, where $\kappa_{1,2}$
are determined by continuity of the pressure. Ref.~\cite{poly} has
shown that such an EOS with $1.5 < \Gamma_{1,2} < 4.0$ and transition
densities $\rho_{12} \approx (2.0 - 3.5) \rho_0$ can mimic a large set
of neutron star matter EOS. We therefore extend the pressure of
neutron star matter based on chiral EFT in this way, with $1.5 <
\Gamma_{1,2} < 4.5$ and $1.5 < \rho_{12}/\rho_0 < 4.5$, as illustrated
in Fig.~\ref{prho}. The possibility of a phase transition at higher
densities is implicitly taken into account if one regards the
$\Gamma_{1,2}$ values as averages over some density range.

\begin{figure}[t]
\begin{center}
\includegraphics[scale=0.75,clip=]{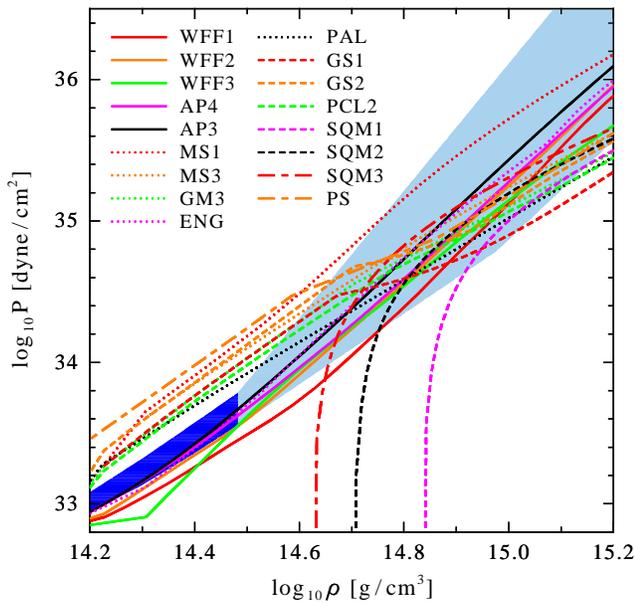}
\end{center}
\vspace*{-4mm}
\caption{Comparison of the EOS based on Fig.~\ref{prho} to a
representative set of EOS used in the literature~\cite{LP}. The
blue band corresponds to the band in Fig.~\ref{prho} and the lighter
region covers the range of polytropes allowed (see text for
discussion).}
\vspace*{-2mm}
\label{prhocomp}
\end{figure}

We solve the TOV equations for the limits of the pressure band below
nuclear densities continued to higher densities by the piecewise
polytropes. The range of $\Gamma_{1,2}$ and $\rho_{12}$ can be
constrained further, first, by causality, which limits the sound speed
to the speed of light
and, second, by the requirement that the EOS support a $1.65 M_\odot$
star~\cite{Freire2009}. The resulting allowed range of polytropes is
shown by the light blue band at higher density in
Fig.~\ref{prhocomp}~\footnote{We have maximized the range of
$\Gamma_1$ given the causality and $1.65 M_\odot$ star
constraint}. The comparison with a representative set of EOS used in
the literature~\cite{LP} demonstrates that the pressure based on
chiral EFT interactions (the darker blue band) sets the scale for the
allowed higher-density extensions and is therefore extremely
important. It also significantly reduces the spread of the pressure at
nuclear densities from a factor $6$ at $\rho_1$ in current neutron
star modeling to a factor $1.5$.

\begin{figure}[t]
\begin{center}
\includegraphics[scale=0.65,clip=]{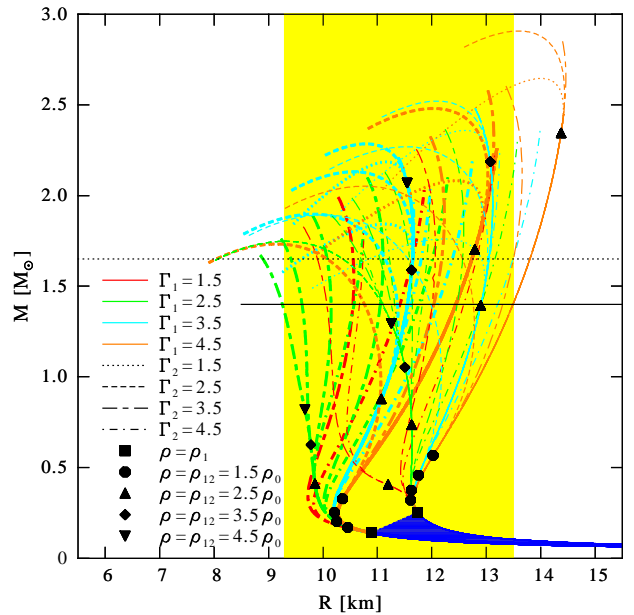}
\vspace*{-4mm}
\end{center}
\caption{Neutron star $M$-$R$ results for the EOS based on
Fig.~\ref{prho}. The thick (thin) lines, corresponding to the
left (right) branch, start from the low pressure limit $c_1 = 
- 0.7 \, {\rm GeV}^{-1}$, $c_3 = - 2.2 \, {\rm GeV}^{-1}$
(high pressure limit $c_1 = - 1.4 \, {\rm GeV}^{-1}$,
$c_3 = - 4.8 \, {\rm GeV}^{-1}$). The blue region corresponds
to the band below nuclear densities in Figs.~\ref{prho} 
and~\ref{prhocomp}. The different piecewise polytropes 
can be identified from the colors/lines indicating
$\Gamma_{1}/\Gamma_{2}$ and from the points denoting $\rho_{12}$. The
vertical band gives the radius constraint for a $1.4 M_\odot$ star.}
\vspace*{-2mm}
\label{MR}
\end{figure}

{\it Results.--} In Fig.~\ref{MR} we show the neutron star $M$-$R$
curves obtained from the allowed EOS range. The blue region
corresponds to the blue band for the pressure in Figs.~\ref{prho}
and~\ref{prhocomp}. At the limits of this region, the pressure of
neutron star matter is continued as piecewise polytropes, and all
curves end when causality is violated. Should causality be violated
before the maximum mass (at $dM/dR=0$) is reached, one could continue
the $M$-$R$ curves by enforcing causality. This would lead to a
somewhat larger maximum mass, but would not affect the masses and
radii of neutron stars with lower central densities. We observe from
the transition density points $\rho_{12}$ in Fig.~\ref{MR} that the
range of $\Gamma_1$ dominates the uncertainty of the general extension
to high densities. Smaller values of $\Gamma_1$ are excluded because
the associated EOS is not able to support a $1.65 M_\odot$ star. The
larger allowed values of the polytropic indices lead to very low
central densities $\rho \sim (2.0 - 2.5) \rho_0$.

We find that the pressure at nuclear densities and below sets the
scale for the $M$-$R$ results. The blue region in Fig.~\ref{MR} ends
almost at the central value of the radius results. For a $1.4 M_\odot$
star, the radius is constrained to $R = 9.3 - 13.5 \, {\rm km}$, as
indicated by the vertical band. While going from neutron matter to
beta equilibrium can reduce individual results for an $1.4 M_\odot$
star by up to $0.4 \km$, the overall result is very similar for pure
neutron matter with $R = 9.3 - 13.3 \, {\rm km}$. Furthermore, if a
$2.0 M_\odot$ star were to be observed, this would reduce the allowed
range to $R = 10.5 - 13.3 \, {\rm km}$.  As for the EOS in
Fig.~\ref{prhocomp}, the presented radius constraint significantly
reduces the spread of viable neutron star models, {\it e.g.}, it is
difficult to see how one can obtain $R \approx 15 \, {\rm km}$ as is
the case for the Shen EOS~\cite{Shen}. Finally, our results are more
rigorous than an estimate based on the empirical $P R^{-4}$
correlation~\cite{LP}, which for the values of the pressure we find,
$P(\rho_0) = 1.4 - 2.1 \mev \fmiq$, implies $R = 9.4 - 11.9 \km$.

\begin{figure}[t]
\begin{center}
\includegraphics[scale=0.65,clip=]{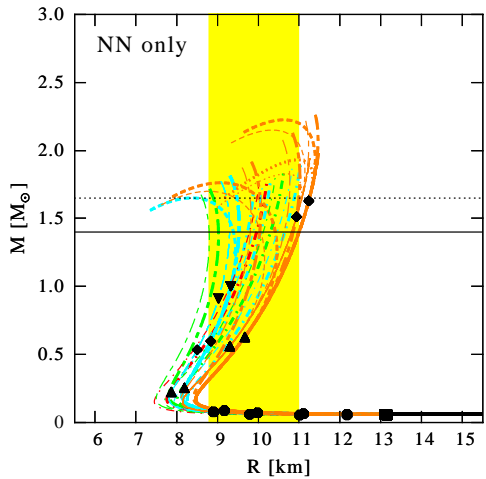}
\end{center}
\vspace*{-4mm}
\caption{Same as Fig.~\ref{MR} but for NN-only interactions (thick
lines are based on EM~\cite{EM} and thin lines on EGM~\cite{EGM}).}
\label{MRNN}
\vspace*{-2mm}
\end{figure}

When chiral 3N forces are neglected, the neutron star radius is
significantly smaller, with $R^{\rm NN} = 8.8 - 11.0 \km$ as shown in
Fig.~\ref{MRNN} based on low-momentum NN interactions only. This
demonstrates that the theoretical error for the radius of a $1.4
M_\odot$ star is due, in about equal amounts, to the
uncertainties in 3N forces and to the extension to higher
densities dominated by $\Gamma_1$.

{\it Effect of the crust.--} In our calculations, the difference
between the neutron and proton masses was neglected and the phases
were assumed to be spatially uniform. In this approximation, matter at
low density consists only of neutrons. The impact of using a more
realistic EOS at low densities can be investigated by observing that
the surface gravity of the star is approximately constant in the outer
layers. By integrating the equation of hydrostatic equilibrium from
the surface of the star up to a crust density $\rho_c$, one finds that
the mass between the density $\rho_c$ and the surface is proportional
to the pressure at $\rho_c$~\cite{Lorenz1993}. Thus the stellar mass
is to a good approximation unaffected by changes in the EOS at $\rho <
\rho_c$. To determine how changes in the low-density EOS affect the
radius, we note that the thickness of the crust ($\rho < \rho_c$) is
$\Delta R=[\mu(\rho_c) - \mu_{\rm s}]/[mg(1+z)]$, where $g=G
M(1+z)/R^2$ is the surface gravity, with surface redshift
$1+z=[1-2GM/(Rc^2)]^{-1/2}$~\cite{Lattimer1994}. Here $\mu_{\rm s}$
is the (neutron) chemical potential at the surface of the star, where
the pressure is zero. For the calculations in this paper, $\mu_{\rm
s}=mc^2$, while for realistic EOS of cold catalysed matter it includes
the binding energy per particle of solid iron, $\approx 8 \mev$. Thus
use of a more realistic EOS at low densities will increase the radius
of the star by $8\, {\rm MeV}/[mg(1+z)]$. This increases the radius
for a $1.4 M_\odot$ star by $0.2 - 0.5 \km$, leading to
our final result $R = 9.7 - 13.9 \, {\rm km}$.

{\it Other implications.--} The relatively weak density dependence of
the nuclear symmetry energy also makes predictions for the neutron
skin of $^{208}$Pb. The symmetry energy of a nucleus in the liquid
droplet model consists largely of bulk and surface contributions, the
latter being determined by an integration of $S_2(\rho)$ through the
nucleus~\cite{Steiner05}. Assuming a quadratic density dependence for
the energy of symmetric nuclear matter and our results for $S_2 =
\overline{S_2} (\rho/\rho_0)^\gamma$, one finds the ratio of the
surface and bulk symmetry parameters to be $S_s/\overline{S}_2 \approx
1.85 \pm 0.25$. This leads to a neutron skin thickness $\delta R =
0.16 - 0.2 \fm$ for $^{208}$Pb, while the correlation with the slope
of the neutron matter energy~\cite{Brown} gives $0.14 - 0.19
\fm$. Therefore we predict $\delta R = 0.14 - 0.2 \fm$, which can be
tested in the parity-violating electron scattering experiment
(PREX)~\cite{Michaels2000}. Finally, we note that in a complementary
approach~\cite{Kurkela2010}, the EOS of high-density matter is
constrained from perturbative QCD calculations, and that our results
are very consistent with the astrophysical estimates of
Ref.~\cite{Steiner2010}.

In this Letter, we have demonstrated that microscopic calculations
based on chiral EFT and many-body theory constrain the pressure of
matter at nuclear densities to within $\pm 25 \%$. This should be
taken into account in modeling stellar collapse, black hole formation,
and neutron stars. Even allowing for uncertainties in the low-energy
theory and the extrapolation to higher densities, we find that the
radius of a neutron star depends only weakly its mass and for a $1.4
M_\odot$ star is rather well constrained.

\begin{acknowledgments}
This paper was initiated at the MICRA2009 Workshop in Copenhagen,
funded in part by the ESF AstroSim and CompStar networks.
We thank the Niels Bohr International Academy for the kind hospitality.
This work was supported in part by NSERC, the U.S.~DOE grant
DE-AC02-87ER40317 and by the Helmholtz Alliance Program of the
Helmholtz Association, contract HA216/EMMI ``Extremes of Density and
Temperature: Cosmic Matter in the Laboratory''. TRIUMF receives
funding via a contribution through the NRC Canada.
\end{acknowledgments}

\end{document}